\documentclass[prb,twocolumn,superscriptaddress,amsmath,amssymb,showpacs,floatfix,preprintnumbers,keywords]{revtex4-2}
\usepackage{amsmath}
\usepackage{amssymb}
\usepackage{graphicx}
\usepackage[export]{adjustbox}
\usepackage{epstopdf}
\usepackage{color}
\usepackage{epsfig}
\usepackage{braket}
\usepackage{amsmath}
\usepackage{stackengine}
\usepackage[colorlinks=true,citecolor=blue,linkcolor=blue]{hyperref}
\usepackage{amssymb}

\DeclareMathAlphabet{\bi}{OML}{cmm}{b}{it}

\begin{document}

\title{Charged vacancy in graphene: interplay between Landau levels and atomic collapse resonances}
\author{Jing Wang}
\email[]{wangjing@hdu.edu.cn}
\affiliation{Key Laboratory of Micro-nano Sensing and IoT of Wenzhou, Wenzhou Institute of Hangzhou Dianzi University, Wenzhou, 325038, China}
\affiliation{School of Electronics and Information, Hangzhou Dianzi University, Hangzhou, Zhejiang Province 310038, China}
\affiliation{Departement Fysica, Universiteit Antwerpen, Groenenborgerlaan 171, B-2020 Antwerpen, Belgium}
\affiliation{NANOlab Center of Excellence, University of Antwerp, Groenenborgerlaan 171, B-2020 Antwerpen, Belgium}

\author{Wen-Sheng Zhao}
\affiliation{School of Electronics and Information, Hangzhou Dianzi University, Hangzhou, Zhejiang Province 310038, China}

\author{Yue Hu}
\affiliation{School of Electronics and Information, Hangzhou Dianzi University, Hangzhou, Zhejiang Province 310038, China}

\author{R. N. Costa Filho}
\affiliation{Departamento de F\'\i sica, Universidade Federal do Cear\'a, Campus do Pici, Fortaleza, Cear\'a, Brazil}

\author{Fran\c{c}ois M. Peeters}
\email[]{francois.peeters@uantwerpen.be}
\affiliation{Departamento de F\'\i sica, Universidade Federal do Cear\'a, Campus do Pici, Fortaleza, Cear\'a, Brazil}
\affiliation{Departement Fysica, Universiteit Antwerpen, Groenenborgerlaan 171, B-2020 Antwerpen, Belgium}
\affiliation{NANOlab Center of Excellence, University of Antwerp, Groenenborgerlaan 171, B-2020 Antwerpen, Belgium}

\begin{abstract}
The interplay between a magnetic field and the Coulomb potential from a charged vacancy on the electron states in graphene is investigated within the tight-binding model. The Coulomb potential removes locally Landau level degeneracy, while the vacancy introduces a satellite level next to the normal Landau level. These satellite levels are found throughout the positive energy region, but in the negative energy region they turn into atomic collapse resonances. Crossings between Landau levels  with different angular quantum number $m$ are found. Unlike the point impurity system in which an anticrossing occurs between Landau levels of the same $m$, in this work anticrossing is found between the normal Landau level and the vacancy induced level. The atomic collapse resonance hybridize with the Landau levels. The charge at which the lowest Landau level $m=-1, N=1$ crosses $E=0$ increases with enhancing magnetic field. Landau level scaling anomaly occurs when the charge is larger than the critical charge $\beta \approx 0.6$ and this critical charge is independent of the magnetic field.
\end{abstract}

\maketitle
\section{Introduction}\label{sec:1}
Ever since the discovery of graphene, it has provided an effective medium to probe analogs and similarities of quantum electrodynamics (QED) phenomena~\cite{ref1}. The charge carriers in graphene are massless Dirac fermions with an effective “speed of light” c$\sim10^6$ m/s~\cite{ref2}. For energies less than about 1 eV the electron spectrum is conical with particular chirality of the  electrons and holes around the high-symmetry K and K’ points. Its unique electric properties allow the detection of the Klein paradox which is a counterintuitive relativistic process~\cite{ref3}. Other QED phenomena, such as anomalous integer quantum Hall effect~\cite{ref4,ref5} and atomic collapse in artifical nuclei was observed on graphene~\cite{ref6}. 

Atomic collapse is a fundamental quantum relativistic phenomenon. It was predicted a century ago but turned out to be impossible to realize in real atoms. By assuming the nucleus to be a point charge, the collapse occurs whenever the charge exceeds the supercritical value $Z>Z_c$ = 137~\cite{ref7,ref8}.  Taking into account the finite size of the nucleus, the condition becomes even more stringent, i.e. $Z_c=170$. However, because of its large effective fine structure constant, the critical charge in graphene is expected to be as low as $Z_c\sim1-2$. By introducing charge impurities with $Z>1$,  atomic collapse has been realized experimentally in several different graphene systems~\cite{ref6,ref9,ref10}. Theoretically, similar phenomena have been intensively studied in both the subcritical and supercritical regimes. Refs.~\cite{ref11,ref12,ref13,ref14,ref15,ref16,ref17,ref18,ref19} studied the atomic collapse in graphene in a single charged impurity field. The extension to the case of two identical impurity charges was considered in Refs.~\cite{ref20,ref21,ref22,ref23,ref24,ref25,ref26}. The interaction between the two impurities splits the atomic collapse state into a pair of bonding and anti-bonding molecular collapse states. Furthermore a new physical regime termed “frustrated supercritical collapse” was demonstrated~\cite{ref25}. When the “artificial nucleus” was realized with a charged vacancy in the graphene lattice weak satellite states appears beside the atomic (molecular) collapse resonances~\cite{ref9,ref26} which are a consequence of the discrete sublattice structure of graphene and the the removal of the equivalence of the two sublattices.  

It has been argued that a strong magnetic field can effectively reduce the value of the critical charge $Z_c$ ~\cite{ref27,ref28,ref29,ref30,ref31}. However, the situation in 2D is different and the effect of a magnetic field on a charged impurity in graphene leads to different conclusions. A theoretical study predicted that the magnetic field drives the critical charge to zero~\cite{ref32}. However, more recent investigations (some focused on an "exact" numerical solution) found that the magnetic field does not affect the value of the critical charge~\cite{ref19,ref33,ref34,ref35,ref36}. Recently, Eren and G$\ddot{u}$\c{c}l$\ddot{u}$ investigated finite size and external magnetic field effects on atomic collapse in a graphene quantum dot and concluded that the size of the quantum dot affects the value of the critical charge \cite{ref37}.

In previous works, a charged impurity was put on top of the graphene layer and it was concluded that: 1) levels with the same orbital number $m$ never cross each other, and 2) an anticrossing occurs between atomic collapse resonance energy levels~\cite{ref19,ref38}. Here we model the "artificial nuclei" by a charged vacancy as in the experiment of Ref.~\cite{ref9} and investigate the effect of a perpendicular magnetic field on the atomic collapse resonance states. In contrast to Refs.~\cite{ref19,ref38} the discrete nature of the graphene sublatttice is retained in our approach and we will use the tight-binding model to calculate the electron states. How these new emerging states influence the crossing between Landau levels and collapse resonances is not clear and therefore triggers our interest and are investigated in this work. Whether the lifting of the sublattice symmetry due to the vacancy will induce any magnetic field dependence of the critical charge for atomic collapse will be critically examined. 

The paper is organized as follows. In section II, we present the model and the method used to obtain the relevant quantities. How do the Landau levels cross atomic collapse resonant states in charged vacancy graphene are studied in section III. The magnetic field dependence of the critical charge for atomic collapse is studied in Section IV. In section V, we summarize our study.

\section{Model}\label{sec:2}
In order to model charged vacancies and to preserve effects due to the discrete lattice (i. e. to go beyond the continuum approach), we use the following tight-binding Hamiltonian  
 \begin{equation}
        \hat{H} = \sum_{\langle i,j\rangle}(t_{ij}  \hat{a}_i^\dagger \hat{b}_j + H.c.) + \sum_i V_i  \hat{a}_i^\dagger \hat{a}_i + \sum_j V_j  \hat{b}_j^\dagger \hat{b}_j,
 \end{equation}
where $\hat{a}_i (\hat{b}_j)$ represents the electron creation operator and $\hat{a}_i^\dagger (\hat{b}_j^\dagger)$ is the annihilation operator of an electron at the sublattice A(B) at site $i$($j$), $t_{ij}=-2.8$ eV is the hopping strength between the nearest neighbors. The last two terms take into account the electrostatic potential $V_{i(j)}$ felt by the electron from the charged vacancy at site $i(j)$. The electric potential of the vacancy with effective charge $\beta=Z\alpha$ is $V(r)=-\hbar v_f\beta/r$, where $Z$ is the value of the charge, $\alpha$ is the fine structure constant of graphene, taking into account of its environment, $\hbar$ is Planck constant divided by 2$\pi$ and $v_f$ is the Fermi velocity. In order to simulate the finite size of the vacancy, a cutoff of the electron potential is introduced by replacing $r$ with $r^*=0.5$ nm when $r<0.5$ nm.  The value of $r^*$ was determined in Ref.~\cite{ref9}. The charge on the vacancy can be tuned by firing voltage pulsed from an STM tip on the vacancy~\cite{ref9}. To model a uniform magnetic field, we make use of Peierls' substitution and replace $t_{ij}$ with $t_{ij}e^{{i2\pi}/{\phi_0\int_{i}^{j}\vec{A}_{ij}d\vec{l}}}$ where $\Phi_0=h/e$ is the magnetic quantum, $h$ is the Planck constant and $\vec{A}_{ij}$ is the magnetic vector potential along the path between sites $i$ and $j$. The magnetic field is perpendicular to the graphene plane and the gauge is taken as $\vec{A}(x,y,z)=B(y,0,0)$.

\begin{figure}
\includegraphics[scale=0.6]{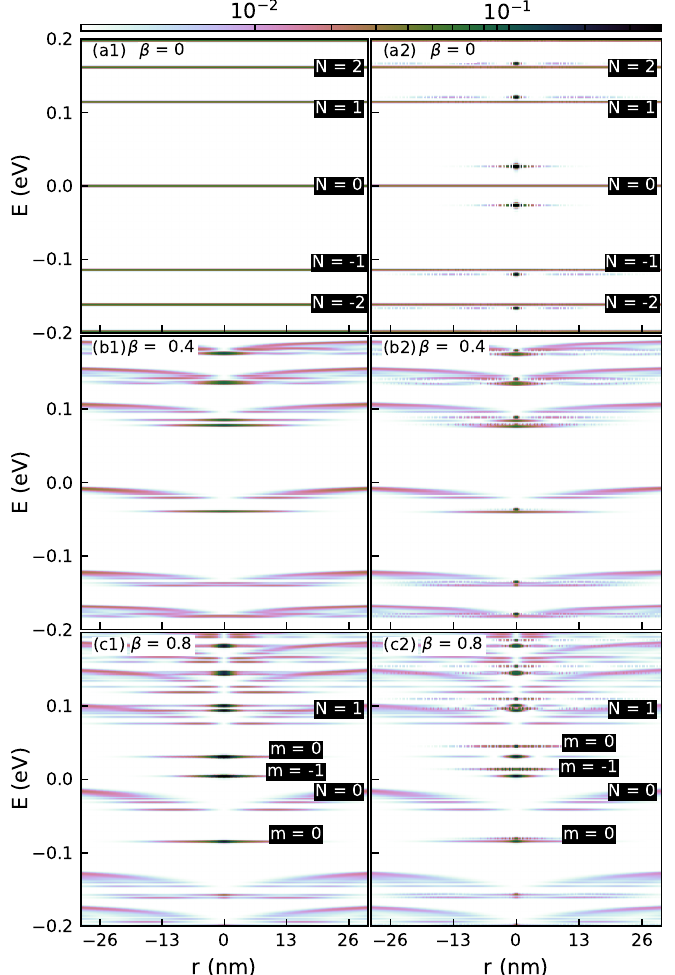}
\caption{\label{fig:fig1} Colormap of the LDOS (in logscale) as a function of position and energy. The results of a charged impurity above graphene (a1-c1) is compared with the results of a charged vacancy (a2-c2). Magnetic field strength is B = 12 T and the vacancy (charged impurity) is located at $r=0$.}
\end{figure}

The eigenvalue problem with the Hamiltonian (1) is solved "numerically exact" on a hexagonal flake with armchair edges to avoid zigzag edges  with zero energy states.  The charged impurity or vacancy are placed in the center of the flake. We take the hexagonal flake edge width of 200 nm which is sufficiently large such that finite size effects are negligible. Such a flake contains more than four million carbon atoms and we use the open source tight-binding Pybinding program to solve the problem numerically~\cite{ref39}. The package employs the kernel polynomial expansion to compute the local density of states (LDOS). An energy broadening of 1 meV is used to simulate effects due to disorder.

\section{Single vacancy: effect of magnetic field}\label{sec:3}
\begin{figure}
\includegraphics[scale=0.5]{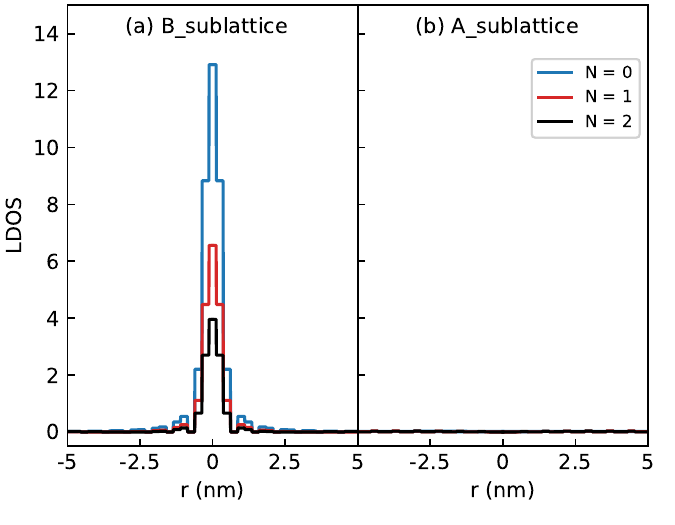}
\caption{\label{fig:fig2} The LDOS as a function of the radial distance $r$ for the vacancy induced level of Landau level $N$ = 0, 1, and 2 in Fig. 1(a2). (a) The LDOS on B sublattice and (b) the LDOS on A sublattice.}
\end{figure}

Firstly, the space-energy map of the electronic states in the subcritical ($0 < \beta < 0.5$) and supercritical ($\beta > 0.5$) regimes are plotted in Fig.~\ref{fig:fig1}. Figs.~\ref{fig:fig1}(a1-c1) are for a point charged impurity which is put 5 nm above graphene and Figs.~\ref{fig:fig1}(a2-c2) are for a charge vacancy system. Fig.~\ref{fig:fig1}(a1) is for pristine graphene and Fig.~\ref{fig:fig1}(a2) for a neutral vacancy in graphene. In the absence of an impurity, Landau levels are independent of the position. Notice that a vacancy introduces a satellite level beside each Landau level. These satellite levels are highly localized around the vacancy and their LDOS intensity rapidly decreases within one nanometer. In order to show the influence of the sublattice, the LDOS of the vacancy induced satellite levels of Landau level $N=$0, 1 and 2 are plotted on different sublattice as a function of the radial distance $r$ in Fig.~\ref{fig:fig2}. The vacancy is formed by removing an A sublattice atom, and we see that these vacancy induced levels are localized on the B sublattice.     

\begin{figure}
\includegraphics[scale=0.6]{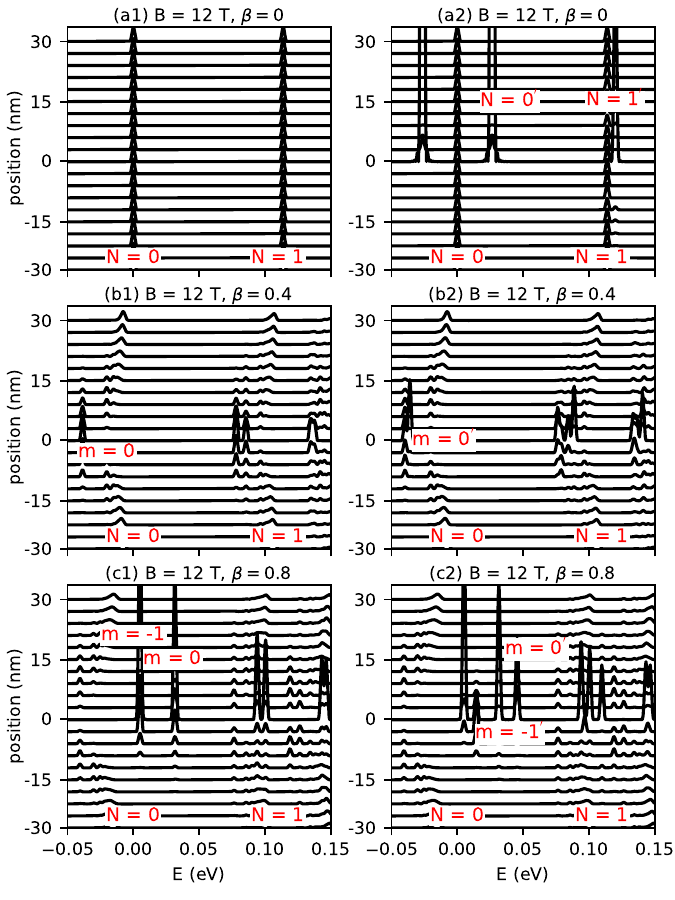}
\caption{\label{fig:fig3} The LDOS as a function of energy for different values of the distance from the charge. (a1-c1) A charged impurity 5 $nm$ above the graphene plane and (a2-c2) for a charged vacancy.}
\end{figure}

Adding charge, the Landau levels start to bend and split into sublevels for different orbital number $m$. Further increasing the charge, many more sublevels appear. In order to show this process clearly, the LDOS as a function of energy for several values of the distance from the impurity is shown in Fig.~\ref{fig:fig3}. The energy region was chosen to include only $N$ = 0 and 1 Landau levels. At $\beta=0.4$ in Fig.~\ref{fig:fig3}(b1), the peak labeled $m=0$ belong to $N$ = 0; at $\beta=0.8$ in Fig.~\ref{fig:fig3}(c1), the two peaks labeled $m=0$ and -1 belong to $N$ = 1. It is clear that the sublevels move down significantly in energy near the charge center. At the position slight away from the charge center, new sublevels are observed but their downward movement is small. When the Landau levels are detected far away from the charge center, they tend not to split and their position is little effected by the charge. By replacing the point charge impurity with a charged vacancy, all of the above properties remain the same except that each level has a satellite level as shown in Figs.~\ref{fig:fig3}(a2-c2). Due to the electron-hole symmetry, the $N$ = 0 Landau level has two vacancy induced satellite levels. In addition, these electron states have a high intensity close to the vacancy and disappear quickly away from it. In the following we show only the results of a charged vacancy system and the LDOS will be computed at the point near the vacancy.

\begin{figure}
\includegraphics[scale=0.5]{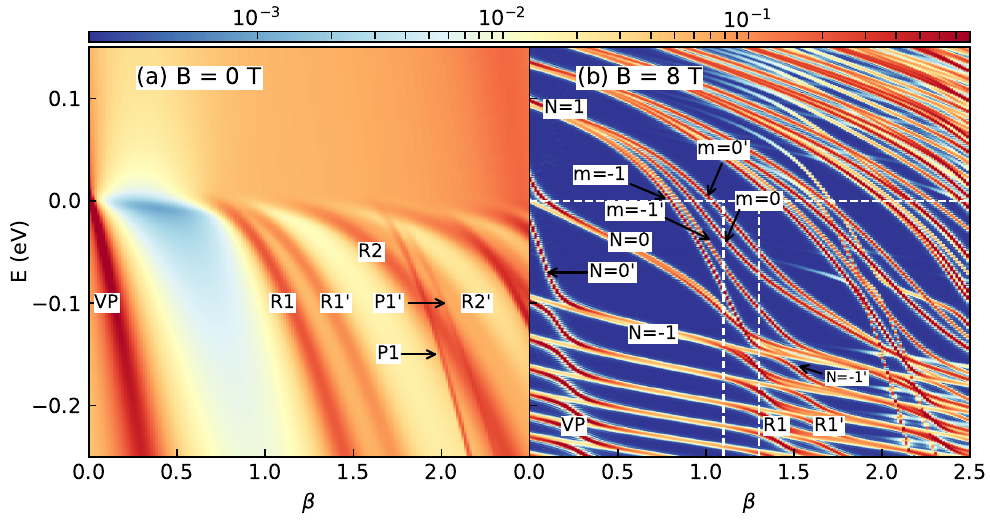}
\caption{\label{fig:fig4}Colormap of the LDOS near the vacancy as a function of energy and charge $\beta$ for (a) B = 0 and (b) B = 8 T. The white dotted vertical lines are (b) at $\beta$ = 1.1 and 1.3. The white dotted horizontal line is at E = 0. The vacancy is formed by removing an A sublattice atom.}
\end{figure}

The LDOS in Fig.~\ref{fig:fig4} is plotted as a function of energy and charge. Without magnetic field, the atomic collapse states and vacancy induced states are recognized from the high LDOS intensity in the negative energy region. The naming of these LDOS resonances are the same as in Ref.~\cite{ref9}. VP represents the vacancy peak. R1 is the 1S atomic collapse state in atoms, R2 is the 2S state and P1 is the 1P state. R1$^{'}$, R2$^{'}$ and P1$^{'}$ are their corresponding vacancy induced satellite states. VP, R1$^{'}$, R2$^{'}$ and P1$^{'}$ are the consequence of the removal of a carbon atom resulting in the breaking of the sublattice symmetry and are absent in the case of a charged impurity system as investigated in Ref.~\cite{ref19} within the continuum approach. 

When a magnetic field is applied, Landau levels are clearly formed at low $\beta$. As the charge increases, Landau levels near the vacancy behave differently in the positive and negative energy region. Landau levels split into individual orbital states with different angular quantum number $m$ in the positive energy region. Crossing and anticrossing are found between Landau levels of different quantum number as Landau levels drop into the negative energy region (e. g. the $N=1$ level splits into  $m=-1,-1',0,0'$ where the accent ($'$) refers to sublattice split level). Refs.~\cite{ref19,ref38} summarized the crossing law as that the level $N=1, m=-1$ crosses level $N=0, m=0$ and is then repelled by level $N=-1,m =-1$ with the formation of an anticrossing following the atomic collapse resonance. This partially remains the same by replacing the point charge impurity with a charged vacancy in this work. Here in addition, the level $N=1, m=-1$ crosses level $N=0, m=0$ but is then repelled by the vacancy induced satellite level $N=-1'$.

\begin{figure}
\includegraphics[scale=0.8]{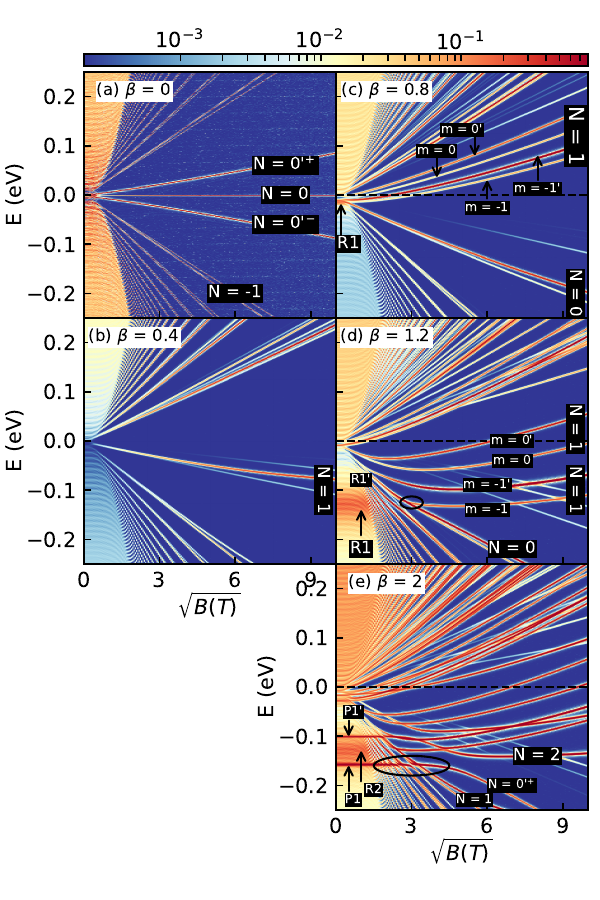}
\caption{\label{fig:fig5}Colormap of LDOS taken near the vacancy as a function of energy and $\sqrt{B}$ for several charges. (a) $\beta$ = 0; (b) $\beta$ = 0.4; (c) $\beta=0.8$; (d) $\beta$ = 1.2 and (e)$\beta$ = 2. The horizontal black dotted line is at E = 0.}
\end{figure}

\begin{figure}
	\includegraphics[scale=0.65]{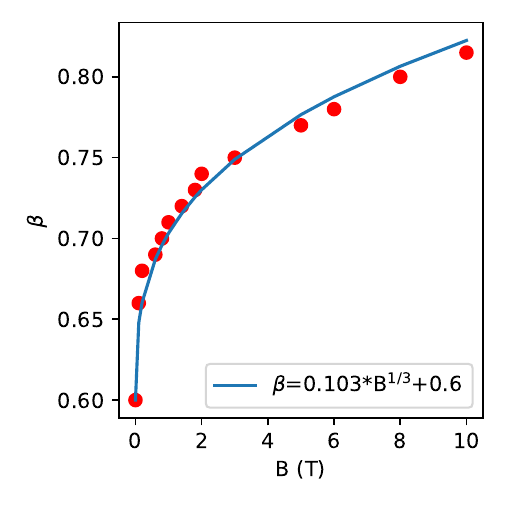}
	\caption{\label{fig:fig6} The charge at which the Landau level $N$ = 1, $m$ = -1 crosses zero energy VS magnetic field. The charges are fitted to $\beta=0.103*B^{1/3}+0.6$.}
\end{figure}


In addition to above phenomenon, Fig.~\ref{fig:fig4}(b) shows some new features due to the broken sublattice symmetry. At small charge,  the value of the LDOS of the vacancy induced electronic state is an order of magnitude larger than the value of the LDOS of normal Landau Levels. It helps us to differentiate vacancy induced levels from normal Landau levels by the color in Fig.~\ref{fig:fig4}(b). The vacancy induced satellite levels (marked with superscript $'$) shift down one energy level through the VP resonance (which is independent of magnetic field) in the negative energy region, e. g., level $N=0^{'}$ moves down to level $N=-1$. As the charge increases, but still less than the critical charge, those parallel levels are normal Landau levels. The inter level spacing of these Landau levels is preserved until landau levels cross the atomic collapse resonance. And the Landau levels shift down one energy level through the atomic collapse resonance. Meanwhile the vacancy induced levels reappear (e. g., level $N=-1'$) after Landau levels cross R1 resonance and they merge into the lower Landau level through R1' resonance. This process repeats and higher orbital states become involved with increase of charge and $\vert E \vert$. 

Another interesting feature in Fig.~\ref{fig:fig4}(b) need to be discussed. The vacancy induced satellite levels exist throughout the positive energy region, but in the negative energy region these levels merge into the normal Landau levels in the region where the vacancy induced resonances do not exist. Thus, the separation distance between the normal Landau level and its satellite level first increases, then decreases with increasing charge $\beta$. In the supplementary information (SI) we show the LDOS calculated on both sublattices separately. They exhibit an out-of-phase oscillation which is similar to what was found in Ref.~\cite{ref9}.

\section{Critical charge for atomic collapse}\label{sec:4}
Next, the LDOS of electronic states are investigated as a function of magnetic field and energy. The results are plotted in Fig.~\ref{fig:fig5} for $\beta$ = 0, 0.4, 0.8, 1.2 and 2. Landau levels show $\sqrt{B}$ behaviour when the charge is smaller than some critical charge as shown in Fig.~\ref{fig:fig5}(a) and (b). As the charge increases beyond some critical charge, the lowest Landau level $m=-1, N=1$ crosses E = 0 and no longer shows the $\sqrt{B}$ scaling as shown in Fig.~\ref{fig:fig5}(c). At small $\sqrt{B}$, an apparent feature is the atomic collapse state that is formed. The $m=-1, N=1$ LL crosses $E=0$ and is included in the atomic collapse resonance. Thus the atomic collapse resonance hybridize with the Landau level. But for larger $\sqrt{B}$ in Fig.~\ref{fig:fig5}(c), the $m=-1, N=1$ LL and the atomic collapse resonance locate at different energies. Further increasing the charge, the R1 atomic collapse resonance moves downward and hybridize again with $m=-1, N=1$ LL at larger $\sqrt{B}$ in Fig.~\ref{fig:fig5}(d). Thus as the magnetic field increases, the charge at which the $m=-1, N=1$ LL crosses $E=0$ increases. In Refs.~\cite{ref32,ref37} the crossing of this LL with $E=0$ was used to determine the value of the critical charge. In Fig.~\ref{fig:fig6}, this charge is plotted as a function of magnetic field and fitted to $\beta=103*B^{1/3}+0.6$. According to this criterion the critical charge increases with B.


 According to the discussion of Fig.~\ref{fig:fig4}, we know that the crossing and anticrossing between Landau levels only occur in the atomic collapse resonance region. This is reconfirmed by Figs.~\ref{fig:fig5}(d and e). The crossing between Landau level $N$ = 1, $m$ = -1 and Landau level $N$ = 0, $m$ = 0 is highlighted by the circle in Fig.~\ref{fig:fig5}(d). As the charge increases, the R1 atomic collapse resonance falls to a lower energy meanwhile R2 and P1 atomic collapse appears as shown in Fig.~\ref{fig:fig5}(e). Crossing and anticrossing between higher order Landau levels are pointed out by the circle in Fig.~\ref{fig:fig5}(e). These crossing and anticrossing points are located at the same energies as the atomic collapse resonances.
 
\begin{figure}
\includegraphics[scale=0.8]{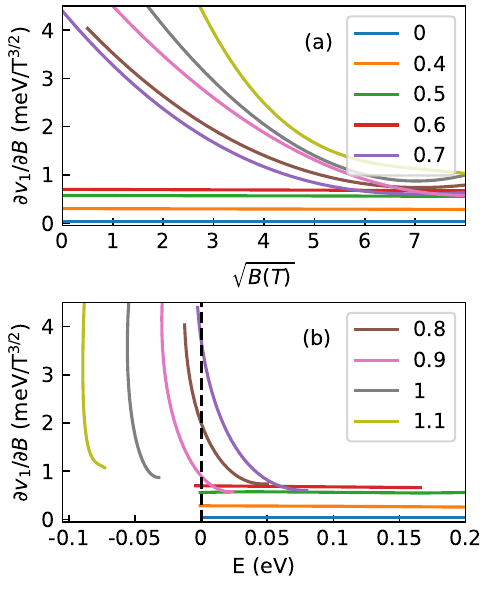}
\caption{\label{fig:fig7} (a) The derivative of the scaling $\nu_1$ for different values of the charge $\beta$ for Landau level $N$ = 1, $m$ = -1. (b) Same derivative, but presented as a function of energy instead of the magnetic field.}
\end{figure}
Previously, the absence of $\sqrt{B}$ scaling of the LL was used to determine the critical charge for atomic collapse~\cite{ref17}.
The energy of the Landau level $N$ can be written as,
\begin{equation}
    E_{N}(B)=\pm{\nu_{F}}\sqrt{2\vert N \vert \hbar}\sqrt{B}=\pm{\nu_{N}}\sqrt{B}
\end{equation}
where $\pm{\nu_{N}}$ is the level scaling prefactor. When $\beta$ is 0, $\partial{\nu_{N}}$/$\partial{B}$ = 0 and Eq.(2) is satisfied. On the other hand, $\partial{\nu_{N}}$/$\partial{B}\neq$ 0 means the level has a scaling anomaly. We use the LDOS data to calculate the derivative $\partial{\nu_{1}}$/$\partial{B}$ for Landau level $N$ = 1, $m$ = -1 and present the results in Fig.~\ref{fig:fig7}. The derivative is almost a constant and close to 0 for $\beta \leqslant$ 0.6, independent of the magnetic field. A constant derivative implies $E_{N}(B)=\pm(\nu_{N}\sqrt{B}+\gamma_{N} B^{3/2})$ with $\gamma$ small and the $\sqrt{B}$ scaling of the LLs is to a large extend satisfied. For $\beta >$ 0.6, there is a significant nonlinear enlargement at small values of the magnetic field. The scaling anomaly is mainly a function of energy and therefore we plot $\partial{\nu_{1}}$/$\partial{B}$ as a function of energy in Fig.~\ref{fig:fig7}(b). Note that for $E<0$, $\partial{\nu_{1}}$/$\partial{B}=0$ when $\beta\leq0.6$ and we have perfect $\sqrt{B}$ scaling. Once $\beta$ increases beyond 0.6, the derivative is non-zero also for $E<0$. Ref.~\cite{ref19} point out that the $\sqrt{B}$ scaling argument is a much better criterion to determine the appearance of the collapse resonance. Without magnetic field, the critical charge in a single impurity system was previously determined in the continuum limit to be $\approx$ 0.5. We found here a slight larger values of $\approx$ 0.6 but the most import conclusion is that magnetic field does not effect the value of the critical charge.

\section{Conclusion}\label{sec:5} 
In this work, we studied how the electronic states of graphene are modified in the presence of a charged vacancy and a perpendicular magnetic field. A charged vacancy causes Landau levels to split into sublevels with different quantum number $m$ and introduces a satellite level next to each normal Landau level. Crossings and anticrossings are formed between Landau levels of different quantum number and the Landau level repulsion occurs between normal Landau level and vacancy induced level. The atomic collapse resonance hybridize with Landau levels and the magnetic field increases the charge at which the lowest Landau level $m=-1, N=1$ crosses $E=0$. Defining the critical charge in terms of the Landau level scaling anomaly, we conclude that the magnetic field does not change the critical charge. As compared to previous results for the continuum Dirac-Kepler problem we find a slightly large value of $\beta_{c}\approx$ 0.6 for the critical charge for atomic collapse.

\begin{acknowledgments}
This work was supported by the National Natural Science Foundation of China (Grant No. 62004053), Zhejiang Provience Natural Science Foundation of China (Grant No. LY19F040006).
\end{acknowledgments} 

\bibliography{references}

\begin{thebibliography}{39}%
\makeatletter
\providecommand \@ifxundefined [1]{%
 \@ifx{#1\undefined}
}%
\providecommand \@ifnum [1]{%
 \ifnum #1\expandafter \@firstoftwo
 \else \expandafter \@secondoftwo
 \fi
}%
\providecommand \@ifx [1]{%
 \ifx #1\expandafter \@firstoftwo
 \else \expandafter \@secondoftwo
 \fi
}%
\providecommand \natexlab [1]{#1}%
\providecommand \enquote  [1]{``#1''}%
\providecommand \bibnamefont  [1]{#1}%
\providecommand \bibfnamefont [1]{#1}%
\providecommand \citenamefont [1]{#1}%
\providecommand \href@noop [0]{\@secondoftwo}%
\providecommand \href [0]{\begingroup \@sanitize@url \@href}%
\providecommand \@href[1]{\@@startlink{#1}\@@href}%
\providecommand \@@href[1]{\endgroup#1\@@endlink}%
\providecommand \@sanitize@url [0]{\catcode `\\12\catcode `\$12\catcode
  `\&12\catcode `\#12\catcode `\^12\catcode `\_12\catcode `\%12\relax}%
\providecommand \@@startlink[1]{}%
\providecommand \@@endlink[0]{}%
\providecommand \url  [0]{\begingroup\@sanitize@url \@url }%
\providecommand \@url [1]{\endgroup\@href {#1}{\urlprefix }}%
\providecommand \urlprefix  [0]{URL }%
\providecommand \Eprint [0]{\href }%
\providecommand \doibase [0]{https://doi.org/}%
\providecommand \selectlanguage [0]{\@gobble}%
\providecommand \bibinfo  [0]{\@secondoftwo}%
\providecommand \bibfield  [0]{\@secondoftwo}%
\providecommand \translation [1]{[#1]}%
\providecommand \BibitemOpen [0]{}%
\providecommand \bibitemStop [0]{}%
\providecommand \bibitemNoStop [0]{.\EOS\space}%
\providecommand \EOS [0]{\spacefactor3000\relax}%
\providecommand \BibitemShut  [1]{\csname bibitem#1\endcsname}%
\let\auto@bib@innerbib\@empty
\bibitem [{\citenamefont {Novoselov}\ \emph {et~al.}(2004)\citenamefont
  {Novoselov}, \citenamefont {Geim}, \citenamefont {Morozov}, \citenamefont
  {Jiang}, \citenamefont {Zhang}, \citenamefont {Dubonos}, \citenamefont
  {Grigorieva},\ and\ \citenamefont {A.}}]{ref1}%
  \BibitemOpen
  \bibfield  {author} {\bibinfo {author} {\bibfnamefont {K.~S.}\ \bibnamefont
  {Novoselov}}, \bibinfo {author} {\bibfnamefont {A.~K.}\ \bibnamefont {Geim}},
  \bibinfo {author} {\bibfnamefont {S.~V.}\ \bibnamefont {Morozov}}, \bibinfo
  {author} {\bibfnamefont {D.}~\bibnamefont {Jiang}}, \bibinfo {author}
  {\bibfnamefont {Y.}~\bibnamefont {Zhang}}, \bibinfo {author} {\bibfnamefont
  {S.~V.}\ \bibnamefont {Dubonos}}, \bibinfo {author} {\bibfnamefont {I.~V.}\
  \bibnamefont {Grigorieva}},\ and\ \bibinfo {author} {\bibfnamefont {F.~A.}\
  \bibnamefont {A.}},\ }\href {https://doi.org/10.1126/science.1102896}
  {\bibfield  {journal} {\bibinfo  {journal} {Science}\ }\textbf {\bibinfo
  {volume} {306}},\ \bibinfo {pages} {666} (\bibinfo {year}
  {2004})}\BibitemShut {NoStop}%
\bibitem [{\citenamefont {Castro~Neto}\ \emph {et~al.}(2009)\citenamefont
  {Castro~Neto}, \citenamefont {Guinea}, \citenamefont {Peres}, \citenamefont
  {Novoselov},\ and\ \citenamefont {Geim}}]{ref2}%
  \BibitemOpen
  \bibfield  {author} {\bibinfo {author} {\bibfnamefont {A.~H.}\ \bibnamefont
  {Castro~Neto}}, \bibinfo {author} {\bibfnamefont {F.}~\bibnamefont {Guinea}},
  \bibinfo {author} {\bibfnamefont {N.~M.~R.}\ \bibnamefont {Peres}}, \bibinfo
  {author} {\bibfnamefont {K.~S.}\ \bibnamefont {Novoselov}},\ and\ \bibinfo
  {author} {\bibfnamefont {A.~K.}\ \bibnamefont {Geim}},\ }\href
  {https://doi.org/10.1103/RevModPhys.81.109} {\bibfield  {journal} {\bibinfo
  {journal} {Rev. Mod. Phys.}\ }\textbf {\bibinfo {volume} {81}},\ \bibinfo
  {pages} {109} (\bibinfo {year} {2009})}\BibitemShut {NoStop}%
\bibitem [{\citenamefont {Katsnelson}\ \emph {et~al.}(2006)\citenamefont
  {Katsnelson}, \citenamefont {Novoselov},\ and\ \citenamefont {Geim}}]{ref3}%
  \BibitemOpen
  \bibfield  {author} {\bibinfo {author} {\bibfnamefont {M.~I.}\ \bibnamefont
  {Katsnelson}}, \bibinfo {author} {\bibfnamefont {K.~S.}\ \bibnamefont
  {Novoselov}},\ and\ \bibinfo {author} {\bibfnamefont {A.~K.}\ \bibnamefont
  {Geim}},\ }\href {https://doi.org/10.1038/nphys384} {\bibfield  {journal}
  {\bibinfo  {journal} {Nat. Phys.}\ }\textbf {\bibinfo {volume} {2}},\
  \bibinfo {pages} {620} (\bibinfo {year} {2006})}\BibitemShut {NoStop}%
\bibitem [{\citenamefont {Novoselov}\ \emph {et~al.}(2005)\citenamefont
  {Novoselov}, \citenamefont {Geim}, \citenamefont {V.}, \citenamefont {Jiang},
  \citenamefont {Katsnelson}, \citenamefont {Grigorieva}, \citenamefont
  {Dubonos},\ and\ \citenamefont {Firsov}}]{ref4}%
  \BibitemOpen
  \bibfield  {author} {\bibinfo {author} {\bibfnamefont {K.~S.}\ \bibnamefont
  {Novoselov}}, \bibinfo {author} {\bibfnamefont {A.~K.}\ \bibnamefont {Geim}},
  \bibinfo {author} {\bibfnamefont {M.~S.}\ \bibnamefont {V.}}, \bibinfo
  {author} {\bibfnamefont {D.}~\bibnamefont {Jiang}}, \bibinfo {author}
  {\bibfnamefont {M.~I.}\ \bibnamefont {Katsnelson}}, \bibinfo {author}
  {\bibfnamefont {I.~V.}\ \bibnamefont {Grigorieva}}, \bibinfo {author}
  {\bibfnamefont {S.~V.}\ \bibnamefont {Dubonos}},\ and\ \bibinfo {author}
  {\bibfnamefont {A.~A.}\ \bibnamefont {Firsov}},\ }\href
  {https://doi.org/10.1038/nature04233} {\bibfield  {journal} {\bibinfo
  {journal} {Nature}\ }\textbf {\bibinfo {volume} {438}},\ \bibinfo {pages}
  {197} (\bibinfo {year} {2005})}\BibitemShut {NoStop}%
\bibitem [{\citenamefont {Zhang}\ \emph {et~al.}(2005)\citenamefont {Zhang},
  \citenamefont {Tan}, \citenamefont {Stormer},\ and\ \citenamefont
  {Kim}}]{ref5}%
  \BibitemOpen
  \bibfield  {author} {\bibinfo {author} {\bibfnamefont {Y.}~\bibnamefont
  {Zhang}}, \bibinfo {author} {\bibfnamefont {Y.-W.}\ \bibnamefont {Tan}},
  \bibinfo {author} {\bibfnamefont {H.~L.}\ \bibnamefont {Stormer}},\ and\
  \bibinfo {author} {\bibfnamefont {P.}~\bibnamefont {Kim}},\ }\href
  {https://doi.org/10.1038/nature04235} {\bibfield  {journal} {\bibinfo
  {journal} {Nature}\ }\textbf {\bibinfo {volume} {438}},\ \bibinfo {pages}
  {201} (\bibinfo {year} {2005})}\BibitemShut {NoStop}%
\bibitem [{\citenamefont {Wang}\ \emph {et~al.}(2013)\citenamefont {Wang},
  \citenamefont {Wong}, \citenamefont {Shytov}, \citenamefont {Brar},
  \citenamefont {Choi}, \citenamefont {Wu}, \citenamefont {Tsai}, \citenamefont
  {Regan}, \citenamefont {Zettl}, \citenamefont {Kawakami}, \citenamefont
  {Louie}, \citenamefont {Levitov},\ and\ \citenamefont {Crommie}}]{ref6}%
  \BibitemOpen
  \bibfield  {author} {\bibinfo {author} {\bibfnamefont {Y.}~\bibnamefont
  {Wang}}, \bibinfo {author} {\bibfnamefont {D.}~\bibnamefont {Wong}}, \bibinfo
  {author} {\bibfnamefont {A.~V.}\ \bibnamefont {Shytov}}, \bibinfo {author}
  {\bibfnamefont {V.~W.}\ \bibnamefont {Brar}}, \bibinfo {author}
  {\bibfnamefont {S.}~\bibnamefont {Choi}}, \bibinfo {author} {\bibfnamefont
  {Q.}~\bibnamefont {Wu}}, \bibinfo {author} {\bibfnamefont {H.-Z.}\
  \bibnamefont {Tsai}}, \bibinfo {author} {\bibfnamefont {W.}~\bibnamefont
  {Regan}}, \bibinfo {author} {\bibfnamefont {A.}~\bibnamefont {Zettl}},
  \bibinfo {author} {\bibfnamefont {R.~K.}\ \bibnamefont {Kawakami}}, \bibinfo
  {author} {\bibfnamefont {S.~G.}\ \bibnamefont {Louie}}, \bibinfo {author}
  {\bibfnamefont {L.~S.}\ \bibnamefont {Levitov}},\ and\ \bibinfo {author}
  {\bibfnamefont {M.~F.}\ \bibnamefont {Crommie}},\ }\href
  {https://doi.org/10.1126/science.1234320} {\bibfield  {journal} {\bibinfo
  {journal} {Science}\ }\textbf {\bibinfo {volume} {340}},\ \bibinfo {pages}
  {734} (\bibinfo {year} {2013})}\BibitemShut {NoStop}%
\bibitem [{\citenamefont {Schweppe}\ \emph {et~al.}(1983)\citenamefont
  {Schweppe}, \citenamefont {Gruppe}, \citenamefont {Bethge}, \citenamefont
  {Bokemeyer}, \citenamefont {Cowan}, \citenamefont {Folger}, \citenamefont
  {Greenberg}, \citenamefont {Grein}, \citenamefont {Ito}, \citenamefont
  {Schule}, \citenamefont {Schwalm}, \citenamefont {Stiebing}, \citenamefont
  {Trautmann}, \citenamefont {Vincent},\ and\ \citenamefont
  {Waldschmidt}}]{ref7}%
  \BibitemOpen
  \bibfield  {author} {\bibinfo {author} {\bibfnamefont {J.}~\bibnamefont
  {Schweppe}}, \bibinfo {author} {\bibfnamefont {A.}~\bibnamefont {Gruppe}},
  \bibinfo {author} {\bibfnamefont {K.}~\bibnamefont {Bethge}}, \bibinfo
  {author} {\bibfnamefont {H.}~\bibnamefont {Bokemeyer}}, \bibinfo {author}
  {\bibfnamefont {T.}~\bibnamefont {Cowan}}, \bibinfo {author} {\bibfnamefont
  {H.}~\bibnamefont {Folger}}, \bibinfo {author} {\bibfnamefont {J.~S.}\
  \bibnamefont {Greenberg}}, \bibinfo {author} {\bibfnamefont {H.}~\bibnamefont
  {Grein}}, \bibinfo {author} {\bibfnamefont {S.}~\bibnamefont {Ito}}, \bibinfo
  {author} {\bibfnamefont {R.}~\bibnamefont {Schule}}, \bibinfo {author}
  {\bibfnamefont {D.}~\bibnamefont {Schwalm}}, \bibinfo {author} {\bibfnamefont
  {K.~E.}\ \bibnamefont {Stiebing}}, \bibinfo {author} {\bibfnamefont
  {N.}~\bibnamefont {Trautmann}}, \bibinfo {author} {\bibfnamefont
  {P.}~\bibnamefont {Vincent}},\ and\ \bibinfo {author} {\bibfnamefont
  {M.}~\bibnamefont {Waldschmidt}},\ }\href
  {https://doi.org/10.1103/PhysRevLett.51.2261} {\bibfield  {journal} {\bibinfo
   {journal} {Phys. Rev. Lett.}\ }\textbf {\bibinfo {volume} {51}},\ \bibinfo
  {pages} {2261} (\bibinfo {year} {1983})}\BibitemShut {NoStop}%
\bibitem [{\citenamefont {Cowan}\ \emph {et~al.}(1985)\citenamefont {Cowan},
  \citenamefont {Backe}, \citenamefont {Begemann}, \citenamefont {Bethge},
  \citenamefont {Bokemeyer}, \citenamefont {Folger}, \citenamefont {Greenberg},
  \citenamefont {Grein}, \citenamefont {Gruppe}, \citenamefont {Kido},
  \citenamefont {Kl\"uver}, \citenamefont {Schwalm}, \citenamefont {Schweppe},
  \citenamefont {Stiebing}, \citenamefont {Trautmann},\ and\ \citenamefont
  {Vincent}}]{ref8}%
  \BibitemOpen
  \bibfield  {author} {\bibinfo {author} {\bibfnamefont {T.}~\bibnamefont
  {Cowan}}, \bibinfo {author} {\bibfnamefont {H.}~\bibnamefont {Backe}},
  \bibinfo {author} {\bibfnamefont {M.}~\bibnamefont {Begemann}}, \bibinfo
  {author} {\bibfnamefont {K.}~\bibnamefont {Bethge}}, \bibinfo {author}
  {\bibfnamefont {H.}~\bibnamefont {Bokemeyer}}, \bibinfo {author}
  {\bibfnamefont {H.}~\bibnamefont {Folger}}, \bibinfo {author} {\bibfnamefont
  {J.~S.}\ \bibnamefont {Greenberg}}, \bibinfo {author} {\bibfnamefont
  {H.}~\bibnamefont {Grein}}, \bibinfo {author} {\bibfnamefont
  {A.}~\bibnamefont {Gruppe}}, \bibinfo {author} {\bibfnamefont
  {Y.}~\bibnamefont {Kido}}, \bibinfo {author} {\bibfnamefont {M.}~\bibnamefont
  {Kl\"uver}}, \bibinfo {author} {\bibfnamefont {D.}~\bibnamefont {Schwalm}},
  \bibinfo {author} {\bibfnamefont {J.}~\bibnamefont {Schweppe}}, \bibinfo
  {author} {\bibfnamefont {K.~E.}\ \bibnamefont {Stiebing}}, \bibinfo {author}
  {\bibfnamefont {N.}~\bibnamefont {Trautmann}},\ and\ \bibinfo {author}
  {\bibfnamefont {P.}~\bibnamefont {Vincent}},\ }\href
  {https://doi.org/10.1103/PhysRevLett.54.1761} {\bibfield  {journal} {\bibinfo
   {journal} {Phys. Rev. Lett.}\ }\textbf {\bibinfo {volume} {54}},\ \bibinfo
  {pages} {1761} (\bibinfo {year} {1985})}\BibitemShut {NoStop}%
\bibitem [{\citenamefont {Mao}\ \emph {et~al.}(2016)\citenamefont {Mao},
  \citenamefont {Jiang}, \citenamefont {Moldovan}, \citenamefont {Li},
  \citenamefont {Watanabe}, \citenamefont {Taniguchi}, \citenamefont {Masir},
  \citenamefont {Peeters},\ and\ \citenamefont {Andrei}}]{ref9}%
  \BibitemOpen
  \bibfield  {author} {\bibinfo {author} {\bibfnamefont {J.}~\bibnamefont
  {Mao}}, \bibinfo {author} {\bibfnamefont {Y.}~\bibnamefont {Jiang}}, \bibinfo
  {author} {\bibfnamefont {D.}~\bibnamefont {Moldovan}}, \bibinfo {author}
  {\bibfnamefont {G.}~\bibnamefont {Li}}, \bibinfo {author} {\bibfnamefont
  {K.}~\bibnamefont {Watanabe}}, \bibinfo {author} {\bibfnamefont
  {T.}~\bibnamefont {Taniguchi}}, \bibinfo {author} {\bibfnamefont {M.~R.}\
  \bibnamefont {Masir}}, \bibinfo {author} {\bibfnamefont {F.~M.}\ \bibnamefont
  {Peeters}},\ and\ \bibinfo {author} {\bibfnamefont {E.~Y.}\ \bibnamefont
  {Andrei}},\ }\href {https://doi.org/10.1038/nphys3665} {\bibfield  {journal}
  {\bibinfo  {journal} {Nat. Phys.}\ }\textbf {\bibinfo {volume} {12}},\
  \bibinfo {pages} {545} (\bibinfo {year} {2016})}\BibitemShut {NoStop}%
\bibitem [{\citenamefont {Jiang}\ \emph {et~al.}(2017)\citenamefont {Jiang},
  \citenamefont {Mao}, \citenamefont {Moldovan}, \citenamefont {Masir},
  \citenamefont {Li}, \citenamefont {Watanabe}, \citenamefont {Taniguchi},
  \citenamefont {Peeters},\ and\ \citenamefont {Andrei}}]{ref10}%
  \BibitemOpen
  \bibfield  {author} {\bibinfo {author} {\bibfnamefont {Y.}~\bibnamefont
  {Jiang}}, \bibinfo {author} {\bibfnamefont {J.}~\bibnamefont {Mao}}, \bibinfo
  {author} {\bibfnamefont {D.}~\bibnamefont {Moldovan}}, \bibinfo {author}
  {\bibfnamefont {M.~R.}\ \bibnamefont {Masir}}, \bibinfo {author}
  {\bibfnamefont {G.}~\bibnamefont {Li}}, \bibinfo {author} {\bibfnamefont
  {K.}~\bibnamefont {Watanabe}}, \bibinfo {author} {\bibfnamefont
  {T.}~\bibnamefont {Taniguchi}}, \bibinfo {author} {\bibfnamefont {F.~M.}\
  \bibnamefont {Peeters}},\ and\ \bibinfo {author} {\bibfnamefont {E.~Y.}\
  \bibnamefont {Andrei}},\ }\href {https://doi.org/10.1038/nnano.2017.181}
  {\bibfield  {journal} {\bibinfo  {journal} {Nat. Nanotechnol.}\ }\textbf
  {\bibinfo {volume} {12}},\ \bibinfo {pages} {1045} (\bibinfo {year}
  {2017})}\BibitemShut {NoStop}%
\bibitem [{\citenamefont {Shytov}\ \emph {et~al.}(2007)\citenamefont {Shytov},
  \citenamefont {Katsnelson},\ and\ \citenamefont {S.}}]{ref11}%
  \BibitemOpen
  \bibfield  {author} {\bibinfo {author} {\bibfnamefont {A.~V.}\ \bibnamefont
  {Shytov}}, \bibinfo {author} {\bibfnamefont {M.~I.}\ \bibnamefont
  {Katsnelson}},\ and\ \bibinfo {author} {\bibfnamefont {L.~L.}\ \bibnamefont
  {S.}},\ }\href {https://doi.org/10.1103/PhysRevLett.99.236801} {\bibfield
  {journal} {\bibinfo  {journal} {Phys. Rev. Lett.}\ }\textbf {\bibinfo
  {volume} {99}},\ \bibinfo {pages} {236801} (\bibinfo {year}
  {2007})}\BibitemShut {NoStop}%
\bibitem [{\citenamefont {Pereira}\ \emph
  {et~al.}(2007{\natexlab{a}})\citenamefont {Pereira}, \citenamefont
  {Nilsson},\ and\ \citenamefont {Neto}}]{ref12}%
  \BibitemOpen
  \bibfield  {author} {\bibinfo {author} {\bibfnamefont {V.~M.}\ \bibnamefont
  {Pereira}}, \bibinfo {author} {\bibfnamefont {J.}~\bibnamefont {Nilsson}},\
  and\ \bibinfo {author} {\bibfnamefont {A.~H.~C.}\ \bibnamefont {Neto}},\
  }\href {https://doi.org/10.1103/PhysRevLett.99.166802} {\bibfield  {journal}
  {\bibinfo  {journal} {Phys. Rev. Lett.}\ }\textbf {\bibinfo {volume} {99}},\
  \bibinfo {pages} {166802} (\bibinfo {year} {2007}{\natexlab{a}})}\BibitemShut
  {NoStop}%
\bibitem [{\citenamefont {Fogler}\ \emph {et~al.}(2007)\citenamefont {Fogler},
  \citenamefont {Novikov},\ and\ \citenamefont {Shklovskii}}]{ref13}%
  \BibitemOpen
  \bibfield  {author} {\bibinfo {author} {\bibfnamefont {M.~M.}\ \bibnamefont
  {Fogler}}, \bibinfo {author} {\bibfnamefont {D.~S.}\ \bibnamefont
  {Novikov}},\ and\ \bibinfo {author} {\bibfnamefont {B.~I.}\ \bibnamefont
  {Shklovskii}},\ }\href {https://doi.org/10.1103/PhysRevB.76.233402}
  {\bibfield  {journal} {\bibinfo  {journal} {Phys. Rev. B}\ }\textbf {\bibinfo
  {volume} {76}},\ \bibinfo {pages} {233402} (\bibinfo {year}
  {2007})}\BibitemShut {NoStop}%
\bibitem [{\citenamefont {Terekhov}\ \emph {et~al.}(2008)\citenamefont
  {Terekhov}, \citenamefont {Milstein}, \citenamefont {Kotov},\ and\
  \citenamefont {Sushkov}}]{ref14}%
  \BibitemOpen
  \bibfield  {author} {\bibinfo {author} {\bibfnamefont {I.~S.}\ \bibnamefont
  {Terekhov}}, \bibinfo {author} {\bibfnamefont {A.~I.}\ \bibnamefont
  {Milstein}}, \bibinfo {author} {\bibfnamefont {V.~N.}\ \bibnamefont
  {Kotov}},\ and\ \bibinfo {author} {\bibfnamefont {O.~P.}\ \bibnamefont
  {Sushkov}},\ }\href {https://doi.org/10.1103/PhysRevLett.100.076803}
  {\bibfield  {journal} {\bibinfo  {journal} {Phys. Rev. Lett.}\ }\textbf
  {\bibinfo {volume} {100}},\ \bibinfo {pages} {076803} (\bibinfo {year}
  {2008})}\BibitemShut {NoStop}%
\bibitem [{\citenamefont {Pereira}\ \emph
  {et~al.}(2007{\natexlab{b}})\citenamefont {Pereira}, \citenamefont
  {Nilsson},\ and\ \citenamefont {Castro~Neto}}]{ref15}%
  \BibitemOpen
  \bibfield  {author} {\bibinfo {author} {\bibfnamefont {V.~M.}\ \bibnamefont
  {Pereira}}, \bibinfo {author} {\bibfnamefont {J.}~\bibnamefont {Nilsson}},\
  and\ \bibinfo {author} {\bibfnamefont {A.~H.}\ \bibnamefont {Castro~Neto}},\
  }\href {https://doi.org/10.1103/PhysRevLett.99.166802} {\bibfield  {journal}
  {\bibinfo  {journal} {Phys. Rev. Lett.}\ }\textbf {\bibinfo {volume} {99}},\
  \bibinfo {pages} {166802} (\bibinfo {year} {2007}{\natexlab{b}})}\BibitemShut
  {NoStop}%
\bibitem [{\citenamefont {Neto}\ \emph {et~al.}(2009)\citenamefont {Neto},
  \citenamefont {Kotov}, \citenamefont {Nilsson}, \citenamefont {Pereira},
  \citenamefont {Peres},\ and\ \citenamefont {Uchoa}}]{ref16}%
  \BibitemOpen
  \bibfield  {author} {\bibinfo {author} {\bibfnamefont {A.~C.}\ \bibnamefont
  {Neto}}, \bibinfo {author} {\bibfnamefont {V.}~\bibnamefont {Kotov}},
  \bibinfo {author} {\bibfnamefont {J.}~\bibnamefont {Nilsson}}, \bibinfo
  {author} {\bibfnamefont {V.}~\bibnamefont {Pereira}}, \bibinfo {author}
  {\bibfnamefont {N.}~\bibnamefont {Peres}},\ and\ \bibinfo {author}
  {\bibfnamefont {B.}~\bibnamefont {Uchoa}},\ }\href
  {https://doi.org/10.1016/j.ssc.2009.02.040} {\bibfield  {journal} {\bibinfo
  {journal} {Solid State Commun.}\ }\textbf {\bibinfo {volume} {149}},\
  \bibinfo {pages} {1094} (\bibinfo {year} {2009})}\BibitemShut {NoStop}%
\bibitem [{\citenamefont {Novikov}(2007)}]{ref17}%
  \BibitemOpen
  \bibfield  {author} {\bibinfo {author} {\bibfnamefont {D.~S.}\ \bibnamefont
  {Novikov}},\ }\href {https://doi.org/10.1103/PhysRevB.76.245435} {\bibfield
  {journal} {\bibinfo  {journal} {Phys. Rev. B}\ }\textbf {\bibinfo {volume}
  {76}},\ \bibinfo {pages} {245435} (\bibinfo {year} {2007})}\BibitemShut
  {NoStop}%
\bibitem [{\citenamefont {Kotov}\ \emph {et~al.}(2012)\citenamefont {Kotov},
  \citenamefont {Uchoa}, \citenamefont {Pereira}, \citenamefont {Guinea},\ and\
  \citenamefont {Castro~Neto}}]{ref18}%
  \BibitemOpen
  \bibfield  {author} {\bibinfo {author} {\bibfnamefont {V.~N.}\ \bibnamefont
  {Kotov}}, \bibinfo {author} {\bibfnamefont {B.}~\bibnamefont {Uchoa}},
  \bibinfo {author} {\bibfnamefont {V.~M.}\ \bibnamefont {Pereira}}, \bibinfo
  {author} {\bibfnamefont {F.}~\bibnamefont {Guinea}},\ and\ \bibinfo {author}
  {\bibfnamefont {A.~H.}\ \bibnamefont {Castro~Neto}},\ }\href
  {https://doi.org/10.1103/RevModPhys.84.1067} {\bibfield  {journal} {\bibinfo
  {journal} {Rev. Mod. Phys.}\ }\textbf {\bibinfo {volume} {84}},\ \bibinfo
  {pages} {1067} (\bibinfo {year} {2012})}\BibitemShut {NoStop}%
\bibitem [{\citenamefont {Moldovan}\ \emph
  {et~al.}(2017{\natexlab{a}})\citenamefont {Moldovan}, \citenamefont {Masir},\
  and\ \citenamefont {Peeters}}]{ref19}%
  \BibitemOpen
  \bibfield  {author} {\bibinfo {author} {\bibfnamefont {D.}~\bibnamefont
  {Moldovan}}, \bibinfo {author} {\bibfnamefont {M.~R.}\ \bibnamefont
  {Masir}},\ and\ \bibinfo {author} {\bibfnamefont {F.~M.}\ \bibnamefont
  {Peeters}},\ }\href {https://doi.org/10.1088/2053-1583/aa9647} {\bibfield
  {journal} {\bibinfo  {journal} {2D Mater.}\ }\textbf {\bibinfo {volume}
  {5}},\ \bibinfo {pages} {015017} (\bibinfo {year}
  {2017}{\natexlab{a}})}\BibitemShut {NoStop}%
\bibitem [{\citenamefont {Pottelberge}\ \emph {et~al.}(2019)\citenamefont
  {Pottelberge}, \citenamefont {Moldovan}, \citenamefont {Milovanovi{\'{c}}},\
  and\ \citenamefont {Peeters}}]{ref20}%
  \BibitemOpen
  \bibfield  {author} {\bibinfo {author} {\bibfnamefont {R.~V.}\ \bibnamefont
  {Pottelberge}}, \bibinfo {author} {\bibfnamefont {D.}~\bibnamefont
  {Moldovan}}, \bibinfo {author} {\bibfnamefont {S.~P.}\ \bibnamefont
  {Milovanovi{\'{c}}}},\ and\ \bibinfo {author} {\bibfnamefont {F.~M.}\
  \bibnamefont {Peeters}},\ }\href {https://doi.org/10.1088/2053-1583/ab3feb}
  {\bibfield  {journal} {\bibinfo  {journal} {2D Mater.}\ }\textbf {\bibinfo
  {volume} {6}},\ \bibinfo {pages} {045047} (\bibinfo {year}
  {2019})}\BibitemShut {NoStop}%
\bibitem [{\citenamefont {De~Martino}\ \emph {et~al.}(2014)\citenamefont
  {De~Martino}, \citenamefont {Kl\"opfer}, \citenamefont {Matrasulov},\ and\
  \citenamefont {Egger}}]{ref21}%
  \BibitemOpen
  \bibfield  {author} {\bibinfo {author} {\bibfnamefont {A.}~\bibnamefont
  {De~Martino}}, \bibinfo {author} {\bibfnamefont {D.}~\bibnamefont
  {Kl\"opfer}}, \bibinfo {author} {\bibfnamefont {D.}~\bibnamefont
  {Matrasulov}},\ and\ \bibinfo {author} {\bibfnamefont {R.}~\bibnamefont
  {Egger}},\ }\href {https://doi.org/10.1103/PhysRevLett.112.186603} {\bibfield
   {journal} {\bibinfo  {journal} {Phys. Rev. Lett.}\ }\textbf {\bibinfo
  {volume} {112}},\ \bibinfo {pages} {186603} (\bibinfo {year}
  {2014})}\BibitemShut {NoStop}%
\bibitem [{\citenamefont {Kl\"{o}pfer}\ \emph {et~al.}(2014)\citenamefont
  {Kl\"{o}pfer}, \citenamefont {Martino}, \citenamefont {Matrasulov},\ and\
  \citenamefont {Egger}}]{ref22}%
  \BibitemOpen
  \bibfield  {author} {\bibinfo {author} {\bibfnamefont {D.}~\bibnamefont
  {Kl\"{o}pfer}}, \bibinfo {author} {\bibfnamefont {A.~D.}\ \bibnamefont
  {Martino}}, \bibinfo {author} {\bibfnamefont {D.~U.}\ \bibnamefont
  {Matrasulov}},\ and\ \bibinfo {author} {\bibfnamefont {R.}~\bibnamefont
  {Egger}},\ }\href {https://doi.org/10.1140/epjb/e2014-50414-8} {\bibfield
  {journal} {\bibinfo  {journal} {Eur. Phys. J. B}\ }\textbf {\bibinfo {volume}
  {87}},\ \bibinfo {pages} {187} (\bibinfo {year} {2014})}\BibitemShut
  {NoStop}%
\bibitem [{\citenamefont {Gorbar}\ \emph {et~al.}(2015)\citenamefont {Gorbar},
  \citenamefont {Gusynin},\ and\ \citenamefont {Sobol}}]{ref23}%
  \BibitemOpen
  \bibfield  {author} {\bibinfo {author} {\bibfnamefont {E.~V.}\ \bibnamefont
  {Gorbar}}, \bibinfo {author} {\bibfnamefont {V.~P.}\ \bibnamefont
  {Gusynin}},\ and\ \bibinfo {author} {\bibfnamefont {O.~O.}\ \bibnamefont
  {Sobol}},\ }\href {https://doi.org/10.1103/PhysRevB.92.235417} {\bibfield
  {journal} {\bibinfo  {journal} {Phys. Rev. B}\ }\textbf {\bibinfo {volume}
  {92}},\ \bibinfo {pages} {235417} (\bibinfo {year} {2015})}\BibitemShut
  {NoStop}%
\bibitem [{\citenamefont {Van~Pottelberge}\ \emph {et~al.}(2018)\citenamefont
  {Van~Pottelberge}, \citenamefont {Van~Duppen},\ and\ \citenamefont
  {Peeters}}]{ref24}%
  \BibitemOpen
  \bibfield  {author} {\bibinfo {author} {\bibfnamefont {R.}~\bibnamefont
  {Van~Pottelberge}}, \bibinfo {author} {\bibfnamefont {B.}~\bibnamefont
  {Van~Duppen}},\ and\ \bibinfo {author} {\bibfnamefont {F.~M.}\ \bibnamefont
  {Peeters}},\ }\href {https://doi.org/10.1103/PhysRevB.98.165420} {\bibfield
  {journal} {\bibinfo  {journal} {Phys. Rev. B}\ }\textbf {\bibinfo {volume}
  {98}},\ \bibinfo {pages} {165420} (\bibinfo {year} {2018})}\BibitemShut
  {NoStop}%
\bibitem [{\citenamefont {Lu}\ \emph {et~al.}(2019)\citenamefont {Lu},
  \citenamefont {Tsai}, \citenamefont {Tatan}, \citenamefont {Wickenburg},
  \citenamefont {Omrani}, \citenamefont {Wong}, \citenamefont {Riss},
  \citenamefont {Piatti}, \citenamefont {Watanabe}, \citenamefont {Taniguchi},
  \citenamefont {Zettl}, \citenamefont {Pereira},\ and\ \citenamefont
  {Crommie}}]{ref25}%
  \BibitemOpen
  \bibfield  {author} {\bibinfo {author} {\bibfnamefont {J.}~\bibnamefont
  {Lu}}, \bibinfo {author} {\bibfnamefont {H.-Z.}\ \bibnamefont {Tsai}},
  \bibinfo {author} {\bibfnamefont {A.~N.}\ \bibnamefont {Tatan}}, \bibinfo
  {author} {\bibfnamefont {S.}~\bibnamefont {Wickenburg}}, \bibinfo {author}
  {\bibfnamefont {A.~A.}\ \bibnamefont {Omrani}}, \bibinfo {author}
  {\bibfnamefont {D.}~\bibnamefont {Wong}}, \bibinfo {author} {\bibfnamefont
  {A.}~\bibnamefont {Riss}}, \bibinfo {author} {\bibfnamefont {E.}~\bibnamefont
  {Piatti}}, \bibinfo {author} {\bibfnamefont {K.}~\bibnamefont {Watanabe}},
  \bibinfo {author} {\bibfnamefont {T.}~\bibnamefont {Taniguchi}}, \bibinfo
  {author} {\bibfnamefont {A.}~\bibnamefont {Zettl}}, \bibinfo {author}
  {\bibfnamefont {V.~M.}\ \bibnamefont {Pereira}},\ and\ \bibinfo {author}
  {\bibfnamefont {M.~F.}\ \bibnamefont {Crommie}},\ }\href
  {https://doi.org/10.1038/s41467-019-08371-2} {\bibfield  {journal} {\bibinfo
  {journal} {Nat. Commun.}\ }\textbf {\bibinfo {volume} {10}},\ \bibinfo
  {pages} {477} (\bibinfo {year} {2019})}\BibitemShut {NoStop}%
\bibitem [{\citenamefont {Wang}\ \emph {et~al.}(2020)\citenamefont {Wang},
  \citenamefont {Anelkovi\'c}, \citenamefont {Wang},\ and\ \citenamefont
  {Peeters}}]{ref26}%
  \BibitemOpen
  \bibfield  {author} {\bibinfo {author} {\bibfnamefont {J.}~\bibnamefont
  {Wang}}, \bibinfo {author} {\bibfnamefont {M.}~\bibnamefont {Anelkovi\'c}},
  \bibinfo {author} {\bibfnamefont {G.}~\bibnamefont {Wang}},\ and\ \bibinfo
  {author} {\bibfnamefont {F.~M.}\ \bibnamefont {Peeters}},\ }\href
  {https://doi.org/10.1103/PhysRevB.102.064108} {\bibfield  {journal} {\bibinfo
   {journal} {Phys. Rev. B}\ }\textbf {\bibinfo {volume} {102}},\ \bibinfo
  {pages} {064108} (\bibinfo {year} {2020})}\BibitemShut {NoStop}%
\bibitem [{\citenamefont {Krainov}\ and\ \citenamefont
  {Zakharov}(1973)}]{ref27}%
  \BibitemOpen
  \bibfield  {author} {\bibinfo {author} {\bibfnamefont {V.~P.}\ \bibnamefont
  {Krainov}}\ and\ \bibinfo {author} {\bibfnamefont {S.~I.}\ \bibnamefont
  {Zakharov}},\ }\bibfield  {title} {\bibinfo {title} {Asymptotic electron
  terms of colliding identical heavy nuclei},\ }\href@noop {} {\bibfield
  {journal} {\bibinfo  {journal} {Sov. Phys.-JETP}\ }\textbf {\bibinfo {volume}
  {37}},\ \bibinfo {pages} {983} (\bibinfo {year} {1973})}\BibitemShut
  {NoStop}%
\bibitem [{\citenamefont {Oraevskii}\ \emph {et~al.}(1977)\citenamefont
  {Oraevskii}, \citenamefont {Rex},\ and\ \citenamefont {Semikoz}}]{ref28}%
  \BibitemOpen
  \bibfield  {author} {\bibinfo {author} {\bibfnamefont {V.~N.}\ \bibnamefont
  {Oraevskii}}, \bibinfo {author} {\bibfnamefont {A.~I.}\ \bibnamefont {Rex}},\
  and\ \bibinfo {author} {\bibfnamefont {V.~B.}\ \bibnamefont {Semikoz}},\
  }\href@noop {} {\bibfield  {journal} {\bibinfo  {journal} {Sov. Phys.-JETP}\
  }\textbf {\bibinfo {volume} {45}},\ \bibinfo {pages} {428} (\bibinfo {year}
  {1977})}\BibitemShut {NoStop}%
\bibitem [{\citenamefont {Karnakov}\ and\ \citenamefont {Popov}(2003)}]{ref29}%
  \BibitemOpen
  \bibfield  {author} {\bibinfo {author} {\bibfnamefont {B.}~\bibnamefont
  {Karnakov}}\ and\ \bibinfo {author} {\bibfnamefont {V.}~\bibnamefont
  {Popov}},\ }\href {https://doi.org/10.1134/1.1633946} {\bibfield  {journal}
  {\bibinfo  {journal} {J. Exp. Theor. Phys.}\ }\textbf {\bibinfo {volume}
  {97}},\ \bibinfo {pages} {890–914} (\bibinfo {year} {2003})}\BibitemShut
  {NoStop}%
\bibitem [{\citenamefont {Vysotskii}\ and\ \citenamefont
  {Godunov}(2014)}]{ref30}%
  \BibitemOpen
  \bibfield  {author} {\bibinfo {author} {\bibfnamefont {M.~I.}\ \bibnamefont
  {Vysotskii}}\ and\ \bibinfo {author} {\bibfnamefont {S.~I.}\ \bibnamefont
  {Godunov}},\ }\href {https://doi.org/10.3367/ufne.0184.201402j.0206}
  {\bibfield  {journal} {\bibinfo  {journal} {Physics-Uspekhi}\ }\textbf
  {\bibinfo {volume} {57}},\ \bibinfo {pages} {194} (\bibinfo {year}
  {2014})}\BibitemShut {NoStop}%
\bibitem [{\citenamefont {Gorbar}\ \emph {et~al.}(2018)\citenamefont {Gorbar},
  \citenamefont {Gusynin},\ and\ \citenamefont {Sobol}}]{ref31}%
  \BibitemOpen
  \bibfield  {author} {\bibinfo {author} {\bibfnamefont {E.~V.}\ \bibnamefont
  {Gorbar}}, \bibinfo {author} {\bibfnamefont {V.~P.}\ \bibnamefont
  {Gusynin}},\ and\ \bibinfo {author} {\bibfnamefont {O.}~\bibnamefont
  {Sobol}},\ }\href {https://doi.org/10.1063/1.5034149} {\bibfield  {journal}
  {\bibinfo  {journal} {Low Temperature Physics}\ }\textbf {\bibinfo {volume}
  {44}},\ \bibinfo {pages} {371} (\bibinfo {year} {2018})}\BibitemShut
  {NoStop}%
\bibitem [{\citenamefont {Gamayun}\ \emph {et~al.}(2011)\citenamefont
  {Gamayun}, \citenamefont {Gorbar},\ and\ \citenamefont {Gusynin}}]{ref32}%
  \BibitemOpen
  \bibfield  {author} {\bibinfo {author} {\bibfnamefont {O.~V.}\ \bibnamefont
  {Gamayun}}, \bibinfo {author} {\bibfnamefont {E.~V.}\ \bibnamefont
  {Gorbar}},\ and\ \bibinfo {author} {\bibfnamefont {V.~P.}\ \bibnamefont
  {Gusynin}},\ }\href {https://doi.org/10.1103/PhysRevB.83.235104} {\bibfield
  {journal} {\bibinfo  {journal} {Phys. Rev. B}\ }\textbf {\bibinfo {volume}
  {83}},\ \bibinfo {pages} {235104} (\bibinfo {year} {2011})}\BibitemShut
  {NoStop}%
\bibitem [{\citenamefont {Valenzuela}\ \emph {et~al.}(2016)\citenamefont
  {Valenzuela}, \citenamefont {Hern{\'{a}}ndez-Ortiz}, \citenamefont {Loewe},\
  and\ \citenamefont {Raya}}]{ref33}%
  \BibitemOpen
  \bibfield  {author} {\bibinfo {author} {\bibfnamefont {D.}~\bibnamefont
  {Valenzuela}}, \bibinfo {author} {\bibfnamefont {S.}~\bibnamefont
  {Hern{\'{a}}ndez-Ortiz}}, \bibinfo {author} {\bibfnamefont {M.}~\bibnamefont
  {Loewe}},\ and\ \bibinfo {author} {\bibfnamefont {A.}~\bibnamefont {Raya}},\
  }\href {https://doi.org/10.1088/1751-8113/49/49/495302} {\bibfield  {journal}
  {\bibinfo  {journal} {J. Phys. A: Math. Theor.}\ }\textbf {\bibinfo {volume}
  {49}},\ \bibinfo {pages} {495302} (\bibinfo {year} {2016})}\BibitemShut
  {NoStop}%
\bibitem [{\citenamefont {Zhang}\ \emph {et~al.}(2012)\citenamefont {Zhang},
  \citenamefont {Barlas},\ and\ \citenamefont {Yang}}]{ref34}%
  \BibitemOpen
  \bibfield  {author} {\bibinfo {author} {\bibfnamefont {Y.}~\bibnamefont
  {Zhang}}, \bibinfo {author} {\bibfnamefont {Y.}~\bibnamefont {Barlas}},\ and\
  \bibinfo {author} {\bibfnamefont {K.}~\bibnamefont {Yang}},\ }\href
  {https://doi.org/10.1103/PhysRevB.85.165423} {\bibfield  {journal} {\bibinfo
  {journal} {Phys. Rev. B}\ }\textbf {\bibinfo {volume} {85}},\ \bibinfo
  {pages} {165423} (\bibinfo {year} {2012})}\BibitemShut {NoStop}%
\bibitem [{\citenamefont {Maiera}\ and\ \citenamefont
  {Siedentopb}(2012)}]{ref35}%
  \BibitemOpen
  \bibfield  {author} {\bibinfo {author} {\bibfnamefont {T.}~\bibnamefont
  {Maiera}}\ and\ \bibinfo {author} {\bibfnamefont {H.}~\bibnamefont
  {Siedentopb}},\ }\href {https://doi.org/10.1063/1.4728982} {\bibfield
  {journal} {\bibinfo  {journal} {J. Math. Phys.}\ }\textbf {\bibinfo {volume}
  {53}},\ \bibinfo {pages} {095207} (\bibinfo {year} {2012})}\BibitemShut
  {NoStop}%
\bibitem [{\citenamefont {Kim}\ and\ \citenamefont {{Eric
  Yang}}(2014)}]{ref36}%
  \BibitemOpen
  \bibfield  {author} {\bibinfo {author} {\bibfnamefont {S.}~\bibnamefont
  {Kim}}\ and\ \bibinfo {author} {\bibfnamefont {S.-R.}\ \bibnamefont {{Eric
  Yang}}},\ }\href {https://doi.org/https://doi.org/10.1016/j.aop.2014.04.022}
  {\bibfield  {journal} {\bibinfo  {journal} {Ann. Phys.}\ }\textbf {\bibinfo
  {volume} {347}},\ \bibinfo {pages} {21} (\bibinfo {year} {2014})}\BibitemShut
  {NoStop}%
\bibitem [{\citenamefont {Eren}\ and\ \citenamefont {Güçlü}(2022)}]{ref37}%
  \BibitemOpen
  \bibfield  {author} {\bibinfo {author} {\bibfnamefont {I.}~\bibnamefont
  {Eren}}\ and\ \bibinfo {author} {\bibfnamefont {A.}~\bibnamefont
  {Güçlü}},\ }\href
  {https://doi.org/https://doi.org/10.1016/j.ssc.2022.114763} {\bibfield
  {journal} {\bibinfo  {journal} {Solid State Communications}\ }\textbf
  {\bibinfo {volume} {351}},\ \bibinfo {pages} {114763} (\bibinfo {year}
  {2022})}\BibitemShut {NoStop}%
\bibitem [{\citenamefont {Sobol}\ \emph {et~al.}(2016)\citenamefont {Sobol},
  \citenamefont {Pyatkovskiy}, \citenamefont {Gorbar},\ and\ \citenamefont
  {Gusynin}}]{ref38}%
  \BibitemOpen
  \bibfield  {author} {\bibinfo {author} {\bibfnamefont {O.~O.}\ \bibnamefont
  {Sobol}}, \bibinfo {author} {\bibfnamefont {P.~K.}\ \bibnamefont
  {Pyatkovskiy}}, \bibinfo {author} {\bibfnamefont {E.~V.}\ \bibnamefont
  {Gorbar}},\ and\ \bibinfo {author} {\bibfnamefont {V.~P.}\ \bibnamefont
  {Gusynin}},\ }\href {https://doi.org/10.1103/PhysRevB.94.115409} {\bibfield
  {journal} {\bibinfo  {journal} {Phys. Rev. B}\ }\textbf {\bibinfo {volume}
  {94}},\ \bibinfo {pages} {115409} (\bibinfo {year} {2016})}\BibitemShut
  {NoStop}%
\bibitem [{\citenamefont {Moldovan}\ \emph
  {et~al.}(2017{\natexlab{b}})\citenamefont {Moldovan}, \citenamefont
  {Andelkovic},\ and\ \citenamefont {Peeters}}]{ref39}%
  \BibitemOpen
  \bibfield  {author} {\bibinfo {author} {\bibfnamefont {D.}~\bibnamefont
  {Moldovan}}, \bibinfo {author} {\bibfnamefont {M.}~\bibnamefont
  {Andelkovic}},\ and\ \bibinfo {author} {\bibfnamefont {F.~M.}\ \bibnamefont
  {Peeters}},\ }\href {https://doi.org/10.5281/zenodo.826942} {\bibinfo {title}
  {Pybinding v0.9.4: A python package for tight-binding calculations}}
  (\bibinfo {year} {2017}{\natexlab{b}})\BibitemShut {NoStop}%
\end{thebibliography}%
\end{document}